\documentclass{aastex62}

\received{...}
\revised{...}
\accepted{...}

\submitjournal{ApJ}

\shorttitle{Reconfiguration and eruption of a filament by reconnection}
\shortauthors{Li et al.}

\begin{document}

\title{Reconfiguration and eruption of a solar filament by magnetic reconnection with an emerging magnetic field}

\correspondingauthor{Leping Li}
\email{lepingli@nao.cas.cn}

\author[0000-0001-5776-056X]{Leping Li}
\affil{National Astronomical Observatories, Chinese Academy of Sciences, Beijing 100101, People's Republic of China}
\affiliation{University of Chinese Academy of Sciences, Beijing 100049, People's Republic of China}

\author[0000-0001-9921-0937]{Hardi Peter}
\author[0000-0002-9270-6785]{Lakshmi Pradeep Chitta}
\affiliation{Max Planck Institute for Solar System Research, 37077 G\"{o}ttingen, Germany}

\author[0000-0001-5705-661X]{Hongqiang Song}
\affiliation{Shandong Provincial Key Laboratory of Optical Astronomy and Solar-Terrestrial Environment, and Institute of Space Sciences, Shandong University, Weihai, Shandong 264209, People's Republic of China}

\author[0000-0002-9121-9686]{Zhe Xu}
\author[0000-0002-5261-6523]{Yongyuan Xiang}
\affiliation{Yunnan Observatories, Chinese Academy of Sciences,  Kunming 650216, People's Republic of China}

\begin{abstract}

Both observations and simulations suggest that the solar filament eruption is closely related to magnetic flux emergence.
It is thought that the eruption is triggered by magnetic reconnection between the filament and the emerging flux.
However, the details of such a reconnection are rarely presented.
In this study, we report the detailed reconnection between a filament and its nearby emerging fields, that led to the reconfiguration and subsequent partial eruption of the filament located over the polarity inversion line of active region 12816. 
Before the reconnection, we observed repeated brightenings in the filament at a location that overlies a site of magnetic flux cancellation. 
Plasmoids form at this brightening region, and propagate bi-directionally along the filament.
These indicate the tether-cutting reconnection that results in the formation and eruption of a flux rope. 
To the northwest of the filament, magnetic fields emerge, and reconnect with the context ones, resulting in repeated jets. 
Afterwards, another magnetic fields emerge near the northwestern filament endpoints, and reconnect with the filament, forming the newly reconnected filament and loops. 
Current sheet repeatedly occurs at the interface, with the mean temperature and emission measure of 1.7\,MK and 1.1$\times$10$^{28}$\,cm$^{-5}$.
Plasmoids form in the current sheet, and propagate along it and further along the newly reconnected filament and loops.
The newly reconnected filament then erupts, while the unreconnected filament remains stable. 
We propose that besides the orientation of emerging fields, some other parameters, such as the position, distance, strength, and area, are also crucial for triggering the filament eruption.

\end{abstract}


\section{Introduction} \label{sec:int}

Solar filaments (prominences) are elongated features, observed as bright prominences above the limb, and dark filaments on the disk due to absorption in chromospheric and coronal diagnostics, e.g., in H$\alpha$ and in extreme ultraviolet (EUV) \citep{2008ApJ...676L..89B, 2010SSRv..151..243L, 2013SoPh..282..147L, 2014LRSP...11....1P, 2018ApJ...863..192L}.
They are composed of cooler and denser chromospheric material suspended in the hotter and rarer corona, and are formed over magnetic polarity inversion lines \citep[PILs;][]{1999A&A...342..867A, 2010SSRv..151..333M, 2015ApJ...806L..13K, 2017ApJ...836L..11S, 2019A&A...627L...5C}.
Magnetic reconnection, rearranging the magnetic field topology, is considered to play an important role in the formation mechanisms of filament \citep{1989ApJ...343..971V, 2017ApJ...845...12K, 2021ApJ...919L..21L}.
Filaments sometimes erupt \citep{2001ApJ...554..474Z, 2006ApJ...651.1238Z, 2018ApJ...869...78C, 2021A&A...653L...2Z}.
Successful filament eruptions are closely associated with other explosive activities on the Sun, such as flares and coronal mass ejections (CMEs) \citep{2012A&A...539A...7L, 2012ApJ...749...12Y, 2013ApJ...765...37F, 2015ApJ...804L..38S, 2018ApJ...860L..25Y}.
These explosive events are believed to be different manifestations of one eruptive process \citep{2001ApJ...559..452Z, 2011SSRv..159...19F, 2012ApJ...748..106T, 2013AdSpR..51.1967S}.

Emergence of magnetic flux is commonly observed in the solar atmosphere. Emergence events with large flux content can lead to the formation of active regions (ARs) \citep{2012A&A...539A...7L}. The early phase of magnetic flux emergence is characterized by the formation of low-lying dark filaments, called arch filament systems (AFS), in the chromosphere \citep{1998SoPh..180..265M, 2018ApJ...855...77S}. In H$\alpha$ observations, blue and red shifts are detected separately at the top and along the legs of AFS, indicating that the cold plasma moves up in the central part and flows down at the footpoints \citep{2002A&A...391..317M}. In this ``leaky bucket" model of AFS formation, during magnetic flux emergence, classical $\Omega$ loops ascend into the atmosphere as the new magnetic flux emerges, and at the same time colder material trapped in the loops drains along both legs \citep{1991A&A...252..353S}. Along with the emergence of magnetic flux, new AFS are formed, while the older ones expand. Overlying these cooler AFS, hot loops are observed at coronal heights, also connecting the emerging magnetic fields \citep{1998SoPh..180..265M, 2002A&A...391..317M}.

Because filaments are progenitors to flares and CMEs, the processes that trigger eruption of a filament need to be well understood.
Various mechanisms triggering the filament eruption have been proposed \citep[e.g.,][]{2004A&A...413L..27T, 2006PhRvL..96y5002K, 2008A&A...492L..35A, 2000ApJ...545..524C}.
By carrying out a statistical study on quiescent filament eruptions, \citet{1995JGR...100.3355F} found that such eruptions usually occur after emergence of magnetic flux in the vicinity of the filaments, with magnetic field orientation that is favorable for reconnection with the preexisting magnetic field.
\citet{1999ApJ...510L.157W} suggested that by diverting the magnetic flux overlying the filament sideways or to greater heights, the new emerging magnetic flux enables the filament eruption. 
This suggestion was then supported by \citet{2007ChJAA...7..129J}, \citet{2007ApJ...669.1359S}, \citet{2018ApJ...862..117D}, and \citet{2019ApJ...874...96Y}.
Conducting a comparative analysis of two filaments in response to the magnetic flux emergence into filament channels, \citet{2015SoPh..290.1687L} proposed that the location of emerging magnetic flux within the filament channel is probably crucial to the filament eruption.
Studying a filament eruption, \citet{2015A&A...583A..47P} suggested that the smooth ascent of the erupting filament is possibly caused by the emerging magnetic flux and torus instability may play a fundamental role, which is helped by the emergence.
Employing H$\alpha$ images from the New Vacuum Solar Telescope \citep[NVST;][]{2014RAA....14..705L}, \citet{2020ApJ...889..106Y} reported that the reconnection between a filament and the emerging magnetic field below the filament triggers the filament eruption.
However, statistically investigating the possible relationship between magnetic field variation and CME initiation, \citet{2008SoPh..250...75Z} noticed that the relationship between CME eruption and magnetic flux emergence is complex, and the appearance of emerging magnetic flux alone is not unique for the CME initiation.

Based on the statistical results of \citet{1995JGR...100.3355F}, \citet{2000ApJ...545..524C} proposed an emerging magnetic flux trigger mechanism for the onset of CME using two-dimensional (2D) magnetohydrodynamic (MHD) numerical simulations of flux ropes, i.e., sheared magnetic fields that support the filament material. 
They found that the reconnection-favored emerging magnetic flux, occurring within the filament channel or on the outer edge of the filament channel, leads to an eruption of the flux rope,
while the non-reconection-favored emerging magnetic flux, happening within the filament channel or on the outer edge of the filament channel, does not trigger a flux rope eruption. 
By including the effects of gravity, spherical geometry in 2.5 dimensions (2.5D), and a stratified ambient medium, \citet{2006A&A...459..927D} extended the emerging magnetic flux trigger mechanism, and obtained similar results.
\citet{2001JGR...10625053L} employed a simple model, and  investigated how an existing magnetic configuration, containing a flux rope, evolves in response to the new emerging magnetic flux.
They noted that there is no simple, universal relation between the orientation of the emerging magnetic flux and the likelihood of the flux rope eruption.
Several other parameters, such as the strength, distance, and area of the emerging magnetic flux, should also be crucial.
Considering three-dimensional (3D) MHD simulations, \citet{2018ApJ...862..117D} modelled the magnetic flux emergence in the vicinity of a flux rope, and suggested that the position of emerging magnetic flux with respect to the background magnetic configuration, as well as other parameters, e.g., the orientation or the amount of emerging magnetic flux, are important for the flux rope eruption.

Many observations and numerical simulations support that reconnection between the emerging magnetic flux and the filament leads to the filament eruption \citep{2000ApJ...545..524C, 2007ApJ...669.1359S, 2018ApJ...862..117D, 2020ApJ...889..106Y}.
However, the details of such reconnection are rarely observed.
In this study, we report the detailed reconnection between a filament and its nearby emerging magnetic field, that results in the reconfiguration and subsequent partial eruption of the filament.
The observations are described in Section\,\ref{sec:obs}. 
The results and a summary and discussion are shown in Sections\,\ref{sec:res} and \ref{sec:sum}, respectively.

\section{Observations}\label{sec:obs}

The Atmospheric Imaging Assembly \citep[AIA;][]{2012SoPh..275...17L} onboard the Solar Dynamic Observatory \citep[SDO;][]{2012SoPh..275....3P} is a set of normal-incidence imaging telescopes. 
It acquires images of the solar atmosphere in 10 wavelength bands.
Different AIA channels show plasma at different temperatures, e.g., 94\,\AA~peaks at $\sim$7.2 MK (Fe XVIII), 335\,\AA~peaks at $\sim$2.5 MK (Fe XVI), 211\,\AA~peaks at $\sim$1.9 MK (Fe XIV), 193\,\AA~peaks at $\sim$1.5 MK (Fe XII), 171\,\AA~peaks at $\sim$0.9 MK (Fe IX), 131\,\AA~peaks at $\sim$0.6 MK (Fe XIII) and $\sim$10 MK (Fe XXI), and 304\,\AA~peaks at $\sim$0.05 MK (He II).
In this study, we employ the AIA 94, 335, 211, 193, 171, 131, and 304\,\AA~images to investigate the evolution of reconnection between the filament and its nearby emerging magnetic field.
These data are recorded at a cadence of 12\,s with spatial sampling of 0.6\arcsec~pixel$^{-1}$.
Here, the AIA EUV images are processed to 1.5 level using ``aia$_{-}$prep.pro". 
The Helioseismic and Magnetic Imager \citep[HMI;][]{2012SoPh..275..229S} on board the SDO provides line-of-sight (LOS) magnetograms, with the time cadence and spatial sampling of 45\,s and 0.5\arcsec~pixel$^{-1}$.
We use the HMI LOS magnetograms to study the evolution of photospheric magnetic fields associated with the reconnection between the emerging magnetic field and the filament.

The NVST, located in the Fuxian Solar Observatory of the Yunnan Observatories, Chinese Academy of Sciences, is a 1\,m ground-based solar telescope.
It provides observations of the solar fine structures and their evolution in the solar lower atmosphere.
On 2021 April 22, the NVST observed the AR 12816 from 05:41\,UT to 09:49\,UT, with a field of view (FOV) of 230\arcsec$\times$230\arcsec~in the H$\alpha$ channel, centered at 6562.8\,\AA~with a bandwidth of 0.25\,\AA.
In this paper, we use the NVST H$\alpha$ images to study the reconnection between the filament and its nearby emerging magnetic field.
The H$\alpha$ images are processed first by flat-field correction and dark current subtraction, and then reconstructed by speckle masking \citep[][and references therein]{2016NewA...49....8X}.
These images have an effective cadence of 42\,s and spatial sampling of 0.164\arcsec~pixel$^{-1}$.
The coalignment of the H$\alpha$ images is carried out by a fast subpixel image registration algorithm \citep{2012JKAS...45..167F, 2015RAA....15..569Y}.
H$\alpha$ images, acquired by GONG instruments operated by NISP/NSO/AURA/NSF with contribution from NOAA, are also employed to check the evolution of the filament.

The NVST H$\alpha$ images have been rotated to match the orientation of SDO observations. All the data from different instruments, e.g., the SDO and NVST, and passbands have been aligned with a principle of best cross-correlation between images of two passbands with the closest characteristic temperatures.

\section{Results}\label{sec:res}

\begin{figure}[ht!]
\centering
\plotone{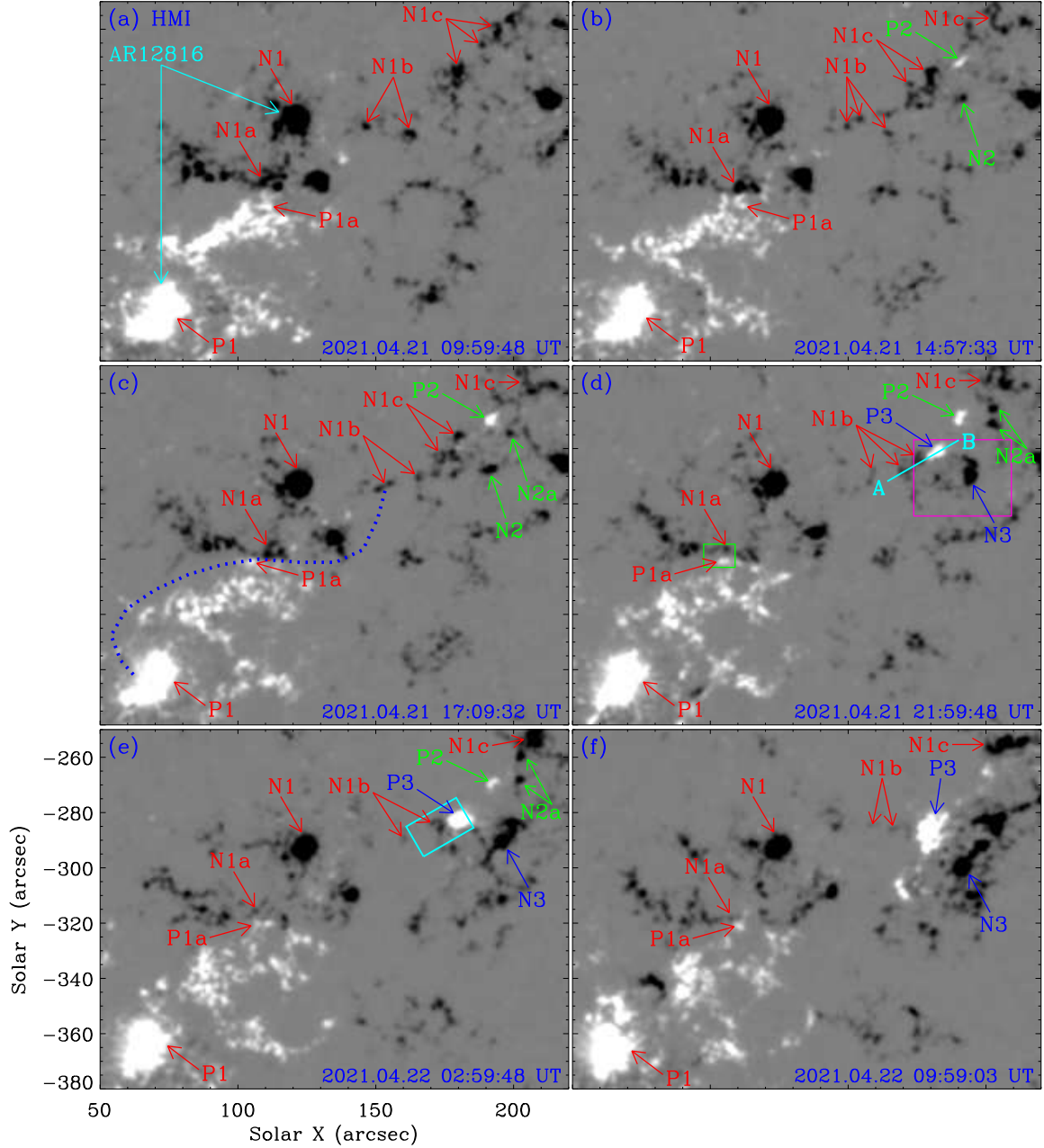}
\caption{Magnetic fields of AR 12816. 
(a)-(f) SDO/HMI LOS magnetograms. 
P1, P1a, P2, and P3, and N1, N1a, N1b, N1c, N2, N2a, and N3 separately indicate the positive and negative magnetic fields.
The blue dotted line in (c) represents the filament in Figure\,\ref{f:filament}(c).
The green, pink, and cyan rectangles in (d) and (e) mark the regions for the magnetic fluxes as shown in Figures\,\ref{f:magnetic_fluxes}(a)-(c), respectively. 
The cyan line AB in (d) shows the position for the time slice of HMI LOS magnetograms displayed in Figure\,\ref{f:magnetic_fluxes}(c). 
An animation of the unannotated HMI LOS magnetograms is available. 
It covers $\sim$1 day starting at 09:45\,UT on 2021 April 21, with a time cadence of 90\,s.
See Section\,\ref{sec:res} and Appendixes A and B for details.
(An animation of this figure is available.) 
\label{f:magnetic_fields}}
\end{figure}

AR 12816 rotated from east to west in the southern hemisphere of the Sun between 2021 April 15 and 27.
In this study, we investigate the evolution of the AR and its associated solar activities during the course of $\sim$1\,day from 09:45\,UT on April 21.
At $\sim$10:00\,UT on April 21, the AR was observed by SDO at the heliographic position S20\,E05. 
It is composed of two main sunspots with positive and negative magnetic fields, labeled by P1 and N1 in Figure\,\ref{f:magnetic_fields}(a), respectively.
Beside these two main sunspots, positive and negative magnetic fields are also identified in the core of this AR, marked by P1a and N1a in Figure\,\ref{f:magnetic_fields}(a).
Scattered negative magnetic fields, denoted by N1b and N1c in Figure\,\ref{f:magnetic_fields}(a), are located to the northwest of these two main sunspots.

\begin{figure}[ht!]
\includegraphics[width=0.78\textwidth]{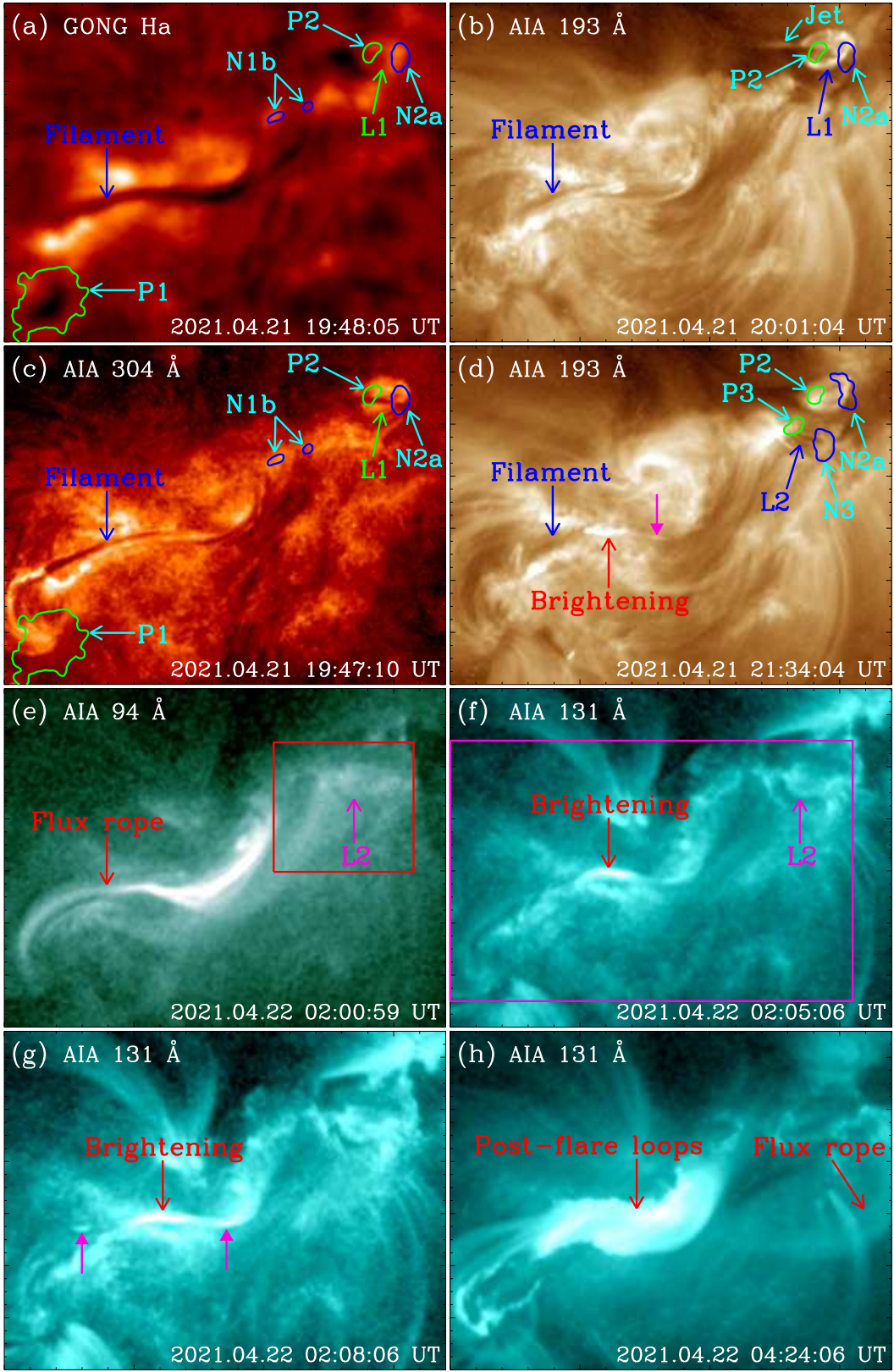}
\centering
\caption{Overview of solar activities in AR 12816.
(a) GONG H$\alpha$, (b) and (d) SDO/AIA 193\,\AA, (c) 304\,\AA, (e) 94\,\AA, and (f)-(h) 131\,\AA~images. 
The green and blue contours in (a)-(d) separately enclose the positive and negative magnetic fields.
L1 and L2 in (a)-(c) and (d)-(f) show the new rising loops separately connecting the emerging magnetic fields P2 and N2a, and P3 and N3. 
The pink solid arrows in (d) and (g) denote the propagating plasmoids.
The red and pink rectangles in (e) and (f) show the FOVs of Figures\,\ref{f:mr_sdo} and \ref{f:mr_nvst}(a), respectively. 
An animation of the unannotated AIA images (panels (c)-(f)) is available.
It covers $\sim$1\,day starting at 09:45\,UT on 2021 April 21, with a time cadence of 2 minutes. 
The FOV is the same as that of Figure\,\ref{f:magnetic_fields}.
See Section\,\ref{sec:res} and Appendixes\,\ref{sec:appa} and \ref{sec:appb} for details.
(An animation of this figure is available.) 
\label{f:filament}}
\end{figure}

In this AR, a sigmoidal filament, with a length of $\sim$120 Mm, is observed above the PIL in both GONG H$\alpha$ and AIA 304\,\AA~images that sample the solar chromosphere; see Figures\,\ref{f:filament}(a) and (c).
It appears to be connected to the magnetic fields P1 and N1b; see the blue dotted line in Figure\,\ref{f:magnetic_fields}(c).
Above the filament, a flux rope is formed in AIA 94 and 131\,\AA~channels by tether-cutting reconnection, constituting a double-decker configuration, and then erupts as a CME; see Appendix A for more details. Furthermore, positive (P2) and negative (N2 and N2a) magnetic fields emerge in the vicinity of the negative polarity patches N1c, resulting in recurring EUV jets by reconnection with the context magnetic fields; see Appendix B for more details. Here, we mainly focus on the reconnection between the filament and its nearby emerging magnetic fields (P3 and N3).

\begin{figure}[ht!]
\plotone{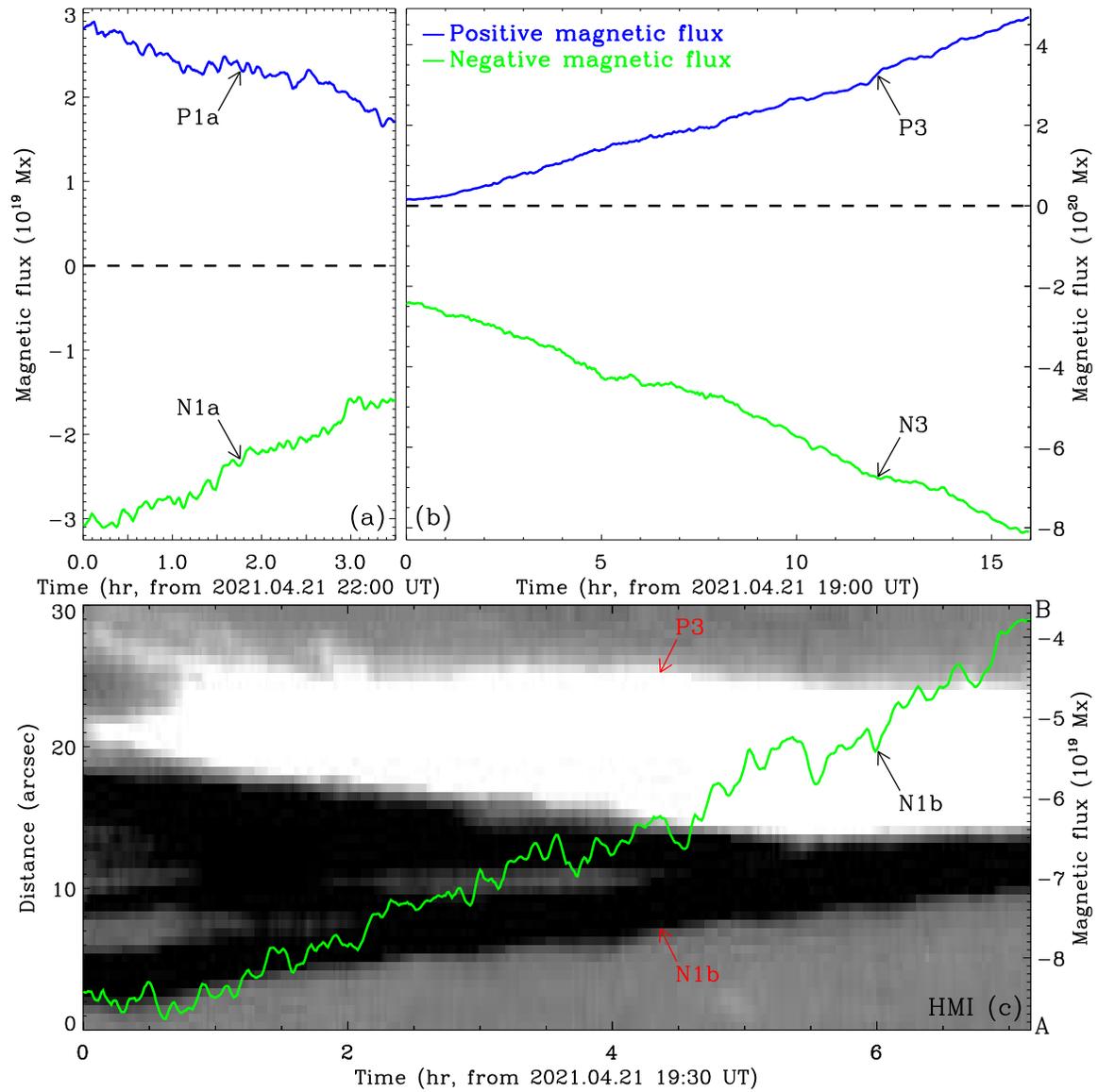}
\centering
\caption{Temporal evolution of magnetic fields in AR 12816.
(a) and (b) magnetic fluxes in the green and pink rectangles in Figure\,\ref{f:magnetic_fields}(d).
(c) Time slice of HMI LOS magnetograms along the cyan line AB in Figure\,\ref{f:magnetic_fields}(d).
The green and blue lines separately show the negative and positive magnetic fluxes.
See Section\,\ref{sec:res} and Appendix A for details.
\label{f:magnetic_fluxes}}
\end{figure}

\begin{figure}[ht!]
\centering
\includegraphics[width=0.6\textwidth]{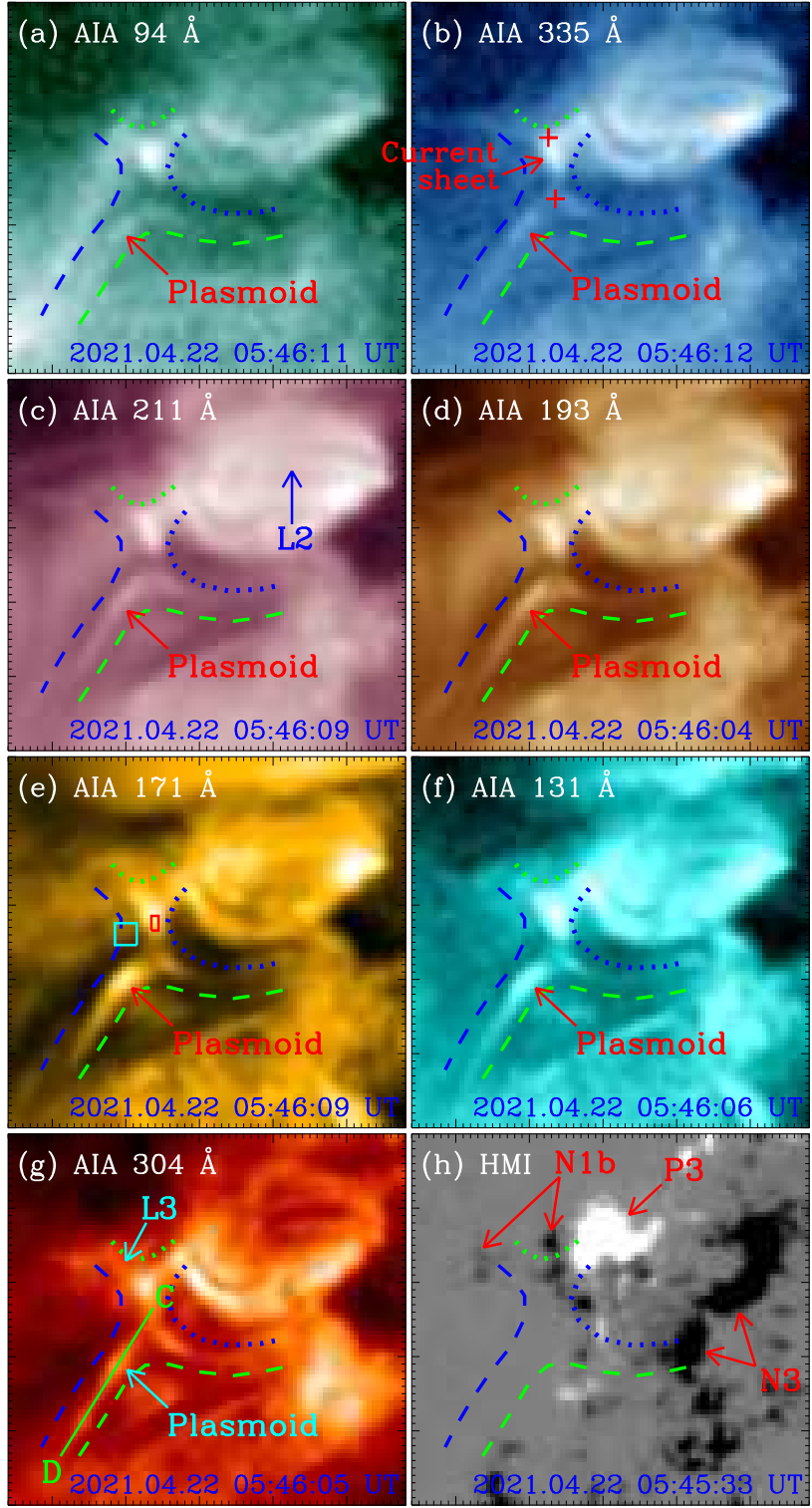}
\caption{Magnetic reconnection between the filament and its nearby emerging loops observed by SDO.
(a) AIA 94\,\AA, (b) 335\,\AA, (c) 211\,\AA, (d) 193\,\AA, (e) 171\,\AA, (f) 131\,\AA, and (g) 304\,\AA~images, and (h) an HMI LOS magnetogram. 
The blue dashed and dotted lines separately outline the reconnecting filament and loops L2.
The green dashed and dotted lines outline the newly reconnected filament and loops L3, respectively.
The red pluses in (b) mark the positions between which the length of the current sheet is measured.
The red and cyan rectangles in (e) separately enclose the region for the DEM curve in Figure\,\ref{f:measurements}(a), and the location where the background emission is computed. 
The green line CD in (g) mark the position for the time slice of AIA 304\,\AA~images as displayed in Figure\,\ref{f:measurements}(b).
An animation of the unannotated SDO observations is available.
It covers $\sim$14\,hr starting at 19:35 UT on 2021 April 21, with a time cadence of 1 minute. 
The FOV is denoted by the red rectangle in Figure\,\ref{f:filament}(e).
See Section\,\ref{sec:res} for details.
(An animation of this figure is available.)
\label{f:mr_sdo}}
\end{figure}

From $\sim$19:20\,UT on April 21, a set of positive and negative magnetic fields P3 and N3 emerged near the negative ones N1b; see Figure\,\ref{f:magnetic_fields}(d). 
The negative magnetic fields N3 and N2 then merged together; see the online animated version of Figure\,\ref{f:magnetic_fields}.  
In the pink rectangle of Figure\,\ref{f:magnetic_fields}(d), we calculate the magnetic fluxes of the emerging magnetic fields P3 and N3, and display them in Figure\,\ref{f:magnetic_fluxes}(b) as blue and green lines.
Magnetic flux of the negative magnetic fields here also includes that of the negative magnetic fields N2.
Hence its magnitude is larger than the positive magnetic flux.
Both the positive and negative magnetic flux increase evidently, with a similar mean increasing rate of $\sim$3$\times$10$^{19}$\,Mx\,hr$^{-1}$.
A set of coronal loops L2 are observed in AIA EUV images, connecting the positive and negative magnetic fields P3 and N3; see Figures\,\ref{f:filament}(d)-(f), and the online animated version of Figure\,\ref{f:filament}.
These new rising loops mostly show emission structures; see Figure\,\ref{f:mr_sdo}, and are seen as dark structures like AFS in the early phase of their evolution; see Figures\,\ref{f:filament}(d)-(f), due to the absorption of the coronal emission by the emerging cool plasma \citep{2012ApJ...754...66P}.

When the emerging loops L2 encounter the nearby filament, X-type structures, outlined by the blue dotted and dashed lines in Figure\,\ref{f:mr_sdo}, form.
At the interface, current sheet is observed in AIA EUV images; see Figure\,\ref{f:mr_sdo} and the online animated version of Figure\,\ref{f:mr_sdo}.
Then we observed the formation of the newly reconnected filament and loops L3, outlined by the green dashed and dotted lines in Figure\,\ref{f:mr_sdo}, suggesting magnetic reconnection.
We measure the length of the current sheet between the two cusp-shaped structures at the ends of the current sheet, marked by red pluses in Figure\,\ref{f:mr_sdo}(b).
The current sheet length ranges in 3.4-14.2\,Mm, with a mean value of 8\,Mm.
For the width of the current sheet, first we get the intensity profile in the AIA 304\,\AA~channel perpendicular to the current sheet.
Using the mean intensity surrounding the current sheet, we compute the background emission, and subtract it from the intensity profile.
We fit the residual intensity profile employing a single Gaussian, and obtain the FWHM of the single Gaussian fit as the current sheet width.
The width of the current sheets ranges from 0.9 to 2.2\,Mm, with a mean value of 1.4\,Mm.
Employing the ratio of the width and the length of the current sheet, we calculate the magnetic reconnection rate, and get a mean value of 0.19 in the range of 0.08-0.41.

The current sheet appears clearly in the AIA EUV channels; see the online animated version of Figure\,\ref{f:mr_sdo}.
Using six AIA EUV channels, including 94, 335, 211, 193, 171, and 131\,\AA, we analyze the temperature and emission measure (EM) of the current sheet.
Here, we employ the differential EM (DEM) analysis using ``xrt\_dem\_iterative2.pro" \citep[][and references therein]{2012ApJ...761...62C}.
The current sheet region, enclosed by the red rectangle in Figure\,\ref{f:mr_sdo}(e), is chosen to compute the DEM.
The region, enclosed by the cyan rectangle in Figure\,\ref{f:mr_sdo}(e), out of the current sheet is chosen for the background emission that is subtracted from the current sheet region.
In each region, the DN counts in each of the six AIA EUV channels are temporally normalized by the exposure time and spatially averaged over all pixels.
The DEM curve of the current sheet region is displayed in Figure\,\ref{f:measurements}(a).
The average DEM-weighted temperature and EM are 1.7\,MK and 1.1$\times$10$^{28}$\,cm$^{-5}$, respectively.
Employing the EM, the electron number density ($n_{e}$) of the current sheet is estimated using $n_{e}$=$\sqrt{\frac{EM}{D}}$, where $D$ is the LOS depth of the current sheet.
Assuming that the depth $D$ equals the width ($W$) of the current sheet, then the electron number density is $n_{e}$=$\sqrt{\frac{EM}{W}}$.
Employing EM=1.1$\times$10$^{28}$\,cm$^{-5}$ and $W$=1.4\,Mm, we obtain the electron number density of the current sheet to be 8.9$\times$10$^{9}$\,cm$^{-3}$.

Bright plasmoids appear in the current sheet in AIA EUV channels, and propagate along it bidirectionally, and further along the newly reconnected filament and loops L3; see Figure\,\ref{f:mr_sdo} and the online animated version of Figure\,\ref{f:mr_sdo}.
Along the green line CD in Figure\,\ref{f:mr_sdo}(g), a time slice of AIA 304\,\AA~images is made, and displayed in Figure\,\ref{f:measurements}(b).
It shows clearly that many plasmoids propagate along the newly reconnected filament, with the moving speeds in the range of 90-240 km\,s$^{-1}$; see the green dotted lines in Figure\,\ref{f:measurements}(b).

At the footpoints of the newly reconnected loops L3, magnetic flux cancellation between the positive and negative magnetic fields P3 and N1b is identified; see Figures\,\ref{f:magnetic_fields}(d)-(f).
Along the cyan line AB in Figure\,\ref{f:magnetic_fields}(d), we make a time slice of HMI LOS magnetograms, and display it in Figure\,\ref{f:magnetic_fluxes}(c).
The positive and negative magnetic fields P3 and N1b move toward each other, collide and cancel with each other; see Figure\,\ref{f:magnetic_fluxes}(c) and the online animated version of Figure\,\ref{f:magnetic_fields}.
In the cyan rectangle in Figure\,\ref{f:magnetic_fields}(e), the magnetic flux of the negative magnetic fields N1b is calculated, and shown in Figure\,\ref{f:magnetic_fluxes}(c) as a green line. 
It decreases evidently, with a mean decreasing rate of $\sim$7$\times$10$^{18}$\,Mx\,hr$^{-1}$.
Here, the magnetic flux of the positive magnetic field P3 is not calculated, because of the simultaneous emergence of the positive and negative magnetic fields P3 and N3; see Figure\,\ref{f:magnetic_fluxes}(b).
Due to the magnetic flux cancellation, some of the negative magnetic fields N1b disappear; see Figure\,\ref{f:magnetic_fields}(f) and the online animated version of Figure\,\ref{f:magnetic_fields}.

\begin{figure}[ht!]
\centering
\includegraphics[width=0.6\textwidth]{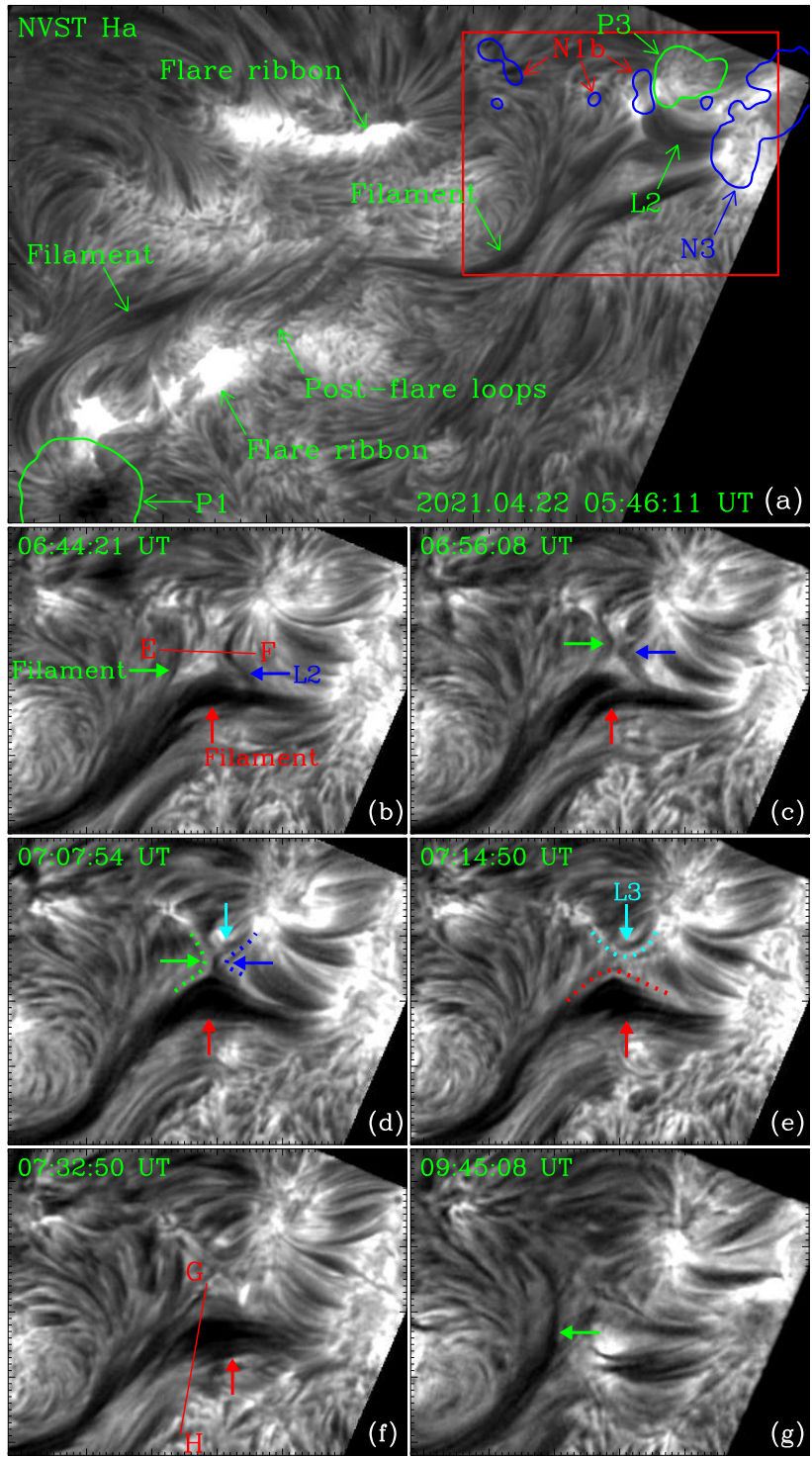}
\caption{Magnetic reconnection between the filament and its nearby emerging loops observed by NVST.
(a)-(g) NVST H$\alpha$ images.
In (a), the red rectangle shows the FOV of (b)-(g), and the green and blue contours separately enclose the positive and negative magnetic fields.
In (b)-(g), the green and blue solid arrows and dotted lines denote the reconnecting filament and loops L2, and the red and cyan solid arrows and dotted lines indicate the newly reconnected filament and loops L3.
The red lines EF and GH in (b) and (f) mark the positions for the time-slices of NVST H$\alpha$ images as shown in Figures\,\ref{f:measurements}(c) and (d).
An animation of the unannotated NVST H$\alpha$ images (panel (b)) is available.
It covers $\sim$4\,hr starting at 05:41\,UT on 2021 April 22, with a time cadence of 42\,s. 
The FOV of (a) is denoted by the pink rectangle in Figure\,\ref{f:filament}(f).
See Section\,\ref{sec:res} for details.
(An animation of this figure is available.)
\label{f:mr_nvst}}
\end{figure}

\begin{figure}[ht!]
\centering
\includegraphics[width=0.9\textwidth]{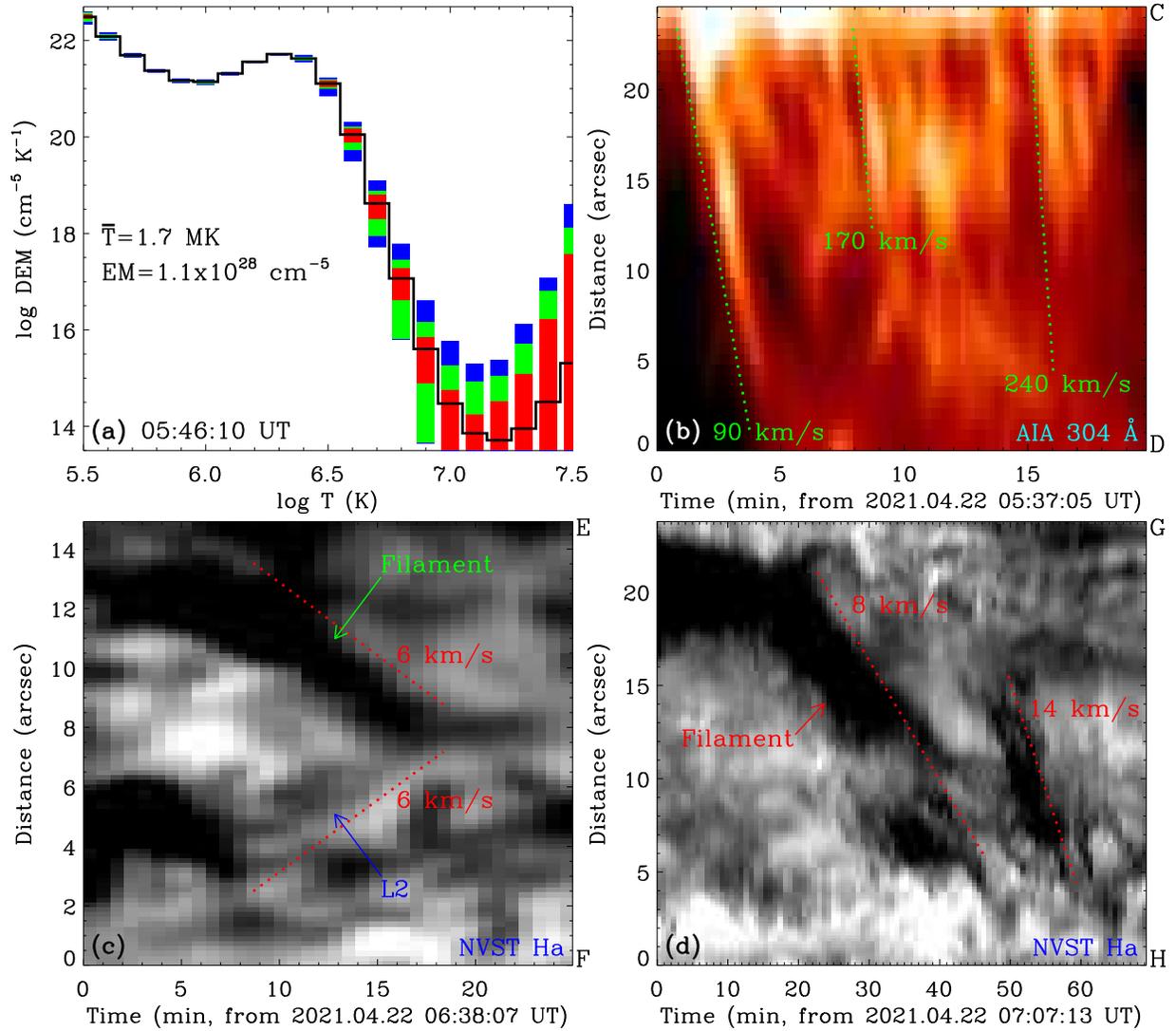}
\caption{Temporal evolution of magnetic reconnection between the filament and its nearby emerging loops.
(a) DEM curve for a current sheet region enclosed by the red rectangle in Figure\,\ref{f:mr_sdo}(e).
(b)-(d) Time-slices of AIA 304\,\AA~and NVST H$\alpha$ images along the green and red lines CD, EF, and GH in Figures\,\ref{f:mr_sdo}(g), \ref{f:mr_nvst}(b), and \ref{f:mr_nvst}(f), respectively.
In (a), the black curve is the best-fit DEM distribution, and the red, green, and blue rectangles separately represent the regions containing 50\%, 51\%-80\%, and 81\%-95\% of the Monte Carlo solutions.
The green and red dotted lines in (b)-(d) outline the motion of plasmoids, the reconnection inflows of filament and loops L2, and the filament eruption, respectively.
See Section\,\ref{sec:res} for details.
\label{f:measurements}}
\end{figure}

The reconnection between the filament and its nearby emerging loops L2 was also observed by the NVST in the chromosphere from 05:41\,UT to 09:49\,UT on April 22.
Figure\,\ref{f:mr_nvst}(a) shows the filament and emerging loops L2 in a NVST H$\alpha$ image.
Different from the AIA EUV observations, in NVST H$\alpha$ images the emerging loops L2 show absorption structures, i.e., AFS; see Figure\,\ref{f:mr_nvst}(a) and the online animated version of Figure\,\ref{f:mr_nvst}.
In this study, the loops L2, i.e., the new rising loops connecting the emerging magnetic fields P3 and N3, thus are separately linked to both the higher-lying hot coronal loops; see Figure\,\ref{f:mr_sdo}, and the lower-lying cold chromospheric AFS; see Figure\,\ref{f:mr_nvst}. They are multi-thermal structures with temperatures ranging from chromospheric to coronal temperatures.
The green and blue contours in Figure\,\ref{f:mr_nvst}(a) enclose the positive and negative magnetic fields P1 and N1b, and P3 and N3, separately linked by the filament and the loops L2.

The filament and loops L2, denoted by the green and blue solid arrows, approach each other; see Figures\,\ref{f:mr_nvst}(b) and (c).
Along the red line EF in Figure\,\ref{f:mr_nvst}(b), a time slice of NVST H$\alpha$ images is made, and displayed in Figure\,\ref{f:measurements}(c).
Obvious inward motions of the filament and loops L2 are identified, with a mean speed of 6\,km\,s$^{-1}$; see the red dotted lines in Figure\,\ref{f:measurements}(c), indicating the reconnection inflows.
Similar to the EUV observations, X-type structures, outlined by the green and blue dotted lines in Figure\,\ref{f:mr_nvst}(d), then form.
At the interface, an elongated sheet-like feature at the same location as the current sheet observed in AIA EUV images, showing absorption feature, takes place; see Figure\,\ref{f:mr_nvst}(d) and the online animated version of Figure\,\ref{f:mr_nvst}. 
Using the similar methods to those employed in the AIA EUV observations, we measure the length, width, and reconnection rate of the sheet-like features in the H$\alpha$ images.
The length (width) of the sheet-like features ranges in 1.5-4.5 (0.4-1.0) Mm, with a mean value of 2.8 (0.7) Mm, and the reconnection rate ranges from 0.11 to 0.6, with a mean value of 0.29.
The sheet-like feature disappears, and the reconnection takes place, with the formation of newly reconnected filament and loops L3, marked by the red and cyan solid arrows and dotted lines in Figure\,\ref{f:mr_nvst}.
Same as the loops L2, the loops L3 in NVST images show dark structures, i.e., AFS. 
They also are multi-thermal structures with temperatures from chromospheric to coronal temperatures, linked to both the higher-lying hot coronal loops; see Figure\,\ref{f:mr_sdo}, and the lower-lying cold chromospheric AFS; see Figure\,\ref{f:mr_nvst}(e), respectively, connecting the positive and negative magnetic fields P3 and N1b.

From $\sim$07:20\,UT on April 22, the newly reconnected filament, denoted by a red solid arrow in Figure\,\ref{f:mr_nvst}(f), erupted; see the online animated version of Figure\,\ref{f:mr_nvst}.
Along the red line GH in Figure\,\ref{f:mr_nvst}(f), we make a time slice of NVST H$\alpha$ images, and show it in Figure\,\ref{f:measurements}(d).
The filament erupts with projection speeds ranging from 8-14 km\,s$^{-1}$; see the red dotted lines in Figure\,\ref{f:measurements}(d). 
The unreconnected filament, marked by a green solid arrow in Figure\,\ref{f:measurements}(g), however, remains stable rather than erupts.
The filament thus only partially erupts.

\section{Summary and discussion}\label{sec:sum}

Employing AIA multi-wavelength images and HMI LOS magnetograms onboard the SDO, in combination with NVST H$\alpha$ images, we investigate the solar activities in AR 12816 in $\sim$1 day from 09:45 UT on 2021 April 21.
The AR mainly consists of positive and negative magnetic fields P1 and N1, between which positive and negative ones P1a and N1a are located.
To the northwest of the AR, negative magnetic fields N1b and N1c are detected.
The lower-lying sigmoidal filament and higher-lying flux rope, forming a double-decker configuration, are observed above the PIL of the AR.
Formation and eruption of a flux rope by reconnection are detected. 
Filament and flux rope formations by tether-cutting reconnection have been suggested theoretically and observationally \citep{2001ApJ...552..833M, 2014ApJ...797L..15C, 2016ApJ...823...41D, 2021ApJ...919L..21L}.
In this study, brightening repeatedly takes place at the middle of the filament.
In the brightening region, plasmoids occur and propagate along the filament bidirectionally.
Furthermore, magnetic flux cancellation between positive and negative magnetic fields P1a and N1a underneath the brightening region is identified.
All these results indicate that tether-cutting reconnection occurs, forming a flux rope, that erupted after the formation, resulting in the associated flare and CME; see Appendix A.
Positive and negative magnetic fields P2, and N2 and N2a, connected by the new rising loops L1, emerge in the context negative ones N1c, with the formation of recurring EUV jets (see Appendix B).
At the jet footpoints, magnetic flux cancellation between the positive and negative magnetic fields P2 and N1c happens.
In the vicinity of the northwestern footpoints of the filament, another positive and negative magnetic fields P3 and N3 emerge, connected by the new rising loops L2.
Magnetic reconnection then occurs between the filament and emerging loops L2, forming the newly reconnected filament and loops L3.
Magnetic flux cancellation between the positive and negative magnetic fields P3 and N1b is detected at the footpoints of newly reconnected loops L3.
The newly reconnected filament then erupts, while the unreconnected filament remains stable, indicating partial eruption of the filament.

\begin{figure}[ht!]
\centering
\includegraphics[width=0.58\textwidth]{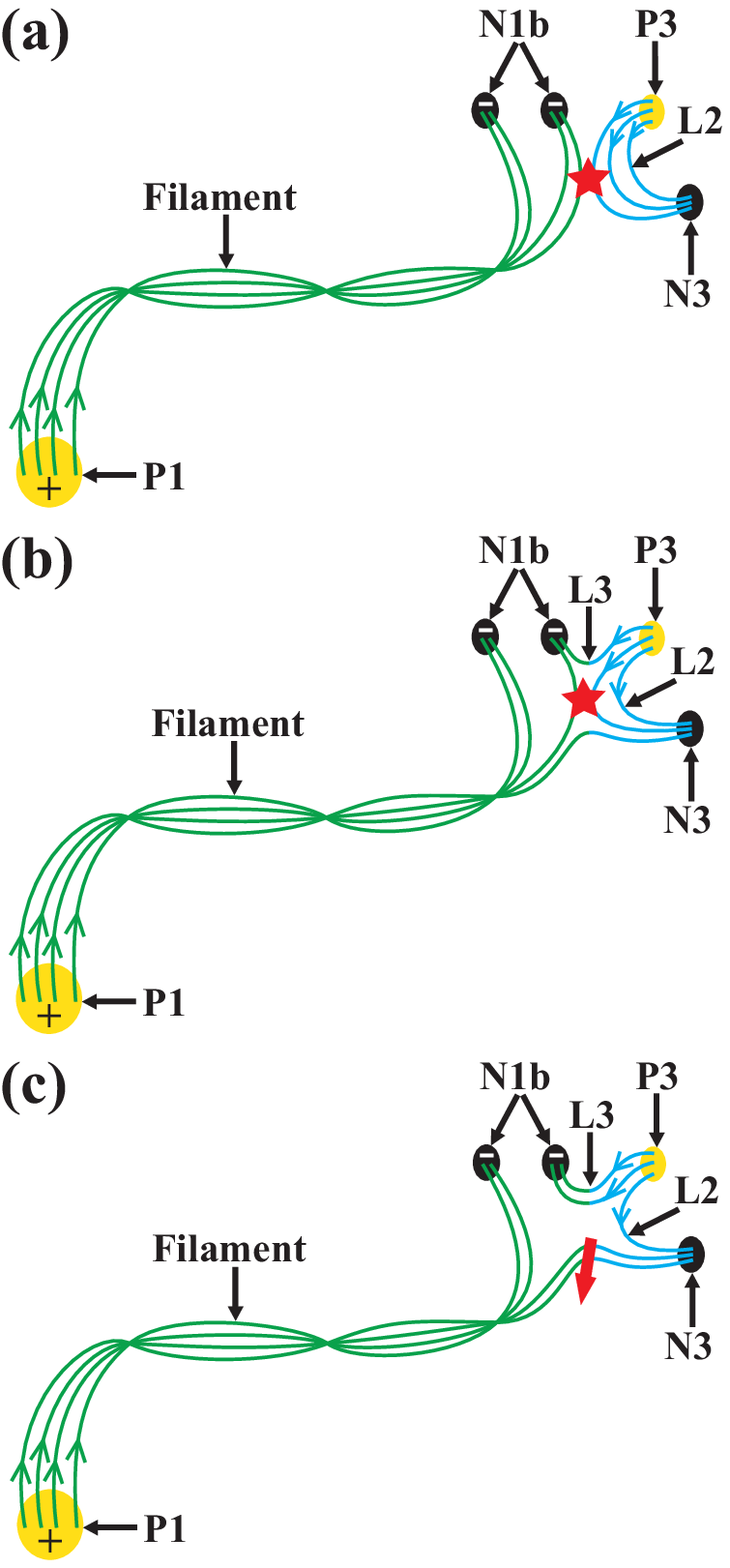}
\caption{Schematic diagrams of the reconfiguration and eruption of the filament by reconnection with the emerging magnetic field.
The yellow (P1 and P3) and black (N1b and N3) ellipses with plus and minus signs represent the positive and negative magnetic fields, respectively.
The green and blue lines indicate magnetic field lines of the filament and emerging loops L2, whose directions are marked by the green and blue arrows. 
The red stars in (a)-(b) denote the magnetic reconnection.
The red solid arrow in (c) indicates the filament eruption.
See Section\,\ref{sec:sum} for details. 
\label{f:cartoon}}
\end{figure}

According to the SDO and NVST observations, schematic diagrams are provided in Figure\,\ref{f:cartoon} to describe the magnetic reconnection between the filament and its nearby emerging magnetic field, and the subsequent eruption of the newly reconnected filament.
Reconnection between the new rising magnetic field lines L2, connecting the emerging positive and negative magnetic fields P3 and N3, and the nearby filament takes place; see the red star between the blue and green lines in Figure\,\ref{f:cartoon}(a).
The newly reconnected magnetic field lines L3 and filament; see the green blue lines in Figure\,\ref{f:cartoon}(b), then form.
The submergence of the newly reconnected magnetic field lines L3; see the green blue lines in Figures\,\ref{f:cartoon}(b)-(c), results in the magnetic flux cancellation during the reconnection process.
The red solid arrow in Figure\,\ref{f:cartoon}(c) denotes the eruption of the newly reconnected filament.

Reconfiguration and eruption of the filament caused by reconnection between the filament and its nearby emerging magnetic fields are observed.
Filament eruptions triggered by reconnection-favored emerging magnetic fields have been suggested observationally and theoretically \citep[e.g.,][]{1995JGR...100.3355F, 2000ApJ...545..524C, 2006A&A...459..927D, 2020ApJ...889..106Y}.
In this study, reconnection between the filament and its nearby emerging loops L2, driven by the emergence of magnetic fields P3 and N3 that are linked by loops L2, forms the newly reconnected filament and loops. 
The newly reconnected filament then erupts due to loss of equilibrium \citep{2000ApJ...545..524C, 2019ApJ...874...96Y}, or driven by the magnetic tension forces \citep{2021NatAs...5.1126J}.
However, most part of the filament, i.e., the unreconnected filament, remains stable, indicating that the filament erupts partially.
These results suggest that only the reconnection-favored orientation of the emerging magnetic fields near the filament may not result in the eruption of the whole filament.
Some other parameters of the emerging magnetic fields, such as the position, distance, strength, and area, are also important for triggering the filament eruption, as suggested by \citet{2001JGR...10625053L}, \citet{2015SoPh..290.1687L}, and \citet{2018ApJ...862..117D}.

The details of reconnection between the filament and its nearby emerging magnetic fields are investigated.
In AIA EUV channels, the width (0.9-2.2\,Mm) and length (3.4-14.2\,Mm) of the current sheets are consistent with those in \citet{2016NatCo...711837X, 2020A&A...633A.121X} and \citet{2021ApJ...908..213L}, but larger than those in \citet{2015ApJ...798L..11Y} and \citet{2018ApJ...858L...4X}.
Moreover, the reconnection rate (0.08-0.41) is identical to those in \citet{2016NatCo...711837X, 2018ApJ...858L...4X} and \citet{2021ApJ...908..213L}, but larger than those in \citet{2020A&A...633A.121X}.
In NVST H$\alpha$ images, similar reconnection rate (0.11-0.6) to that in AIA EUV channels is obtained.
However, the width (0.4-1.0\,Mm) and length (1.5-4.5\,Mm) of the sheet-like features are almost half of those of the current sheets in AIA EUV channels.
This difference may be caused by the different reconnection regions, e.g., in the corona and chromosphere, separately observed by different wavelength channels of the AIA and NVST with different spatial samplings. 
Many plasmoids form in the current sheets in AIA EUV channels, indicating the presence of plasmoid instabilities during the reconnection process \citep{li16a, 2021ApJ...908..213L, 2019A&A...628A...8P}.
The propagating speeds (90-240 km\,s$^{-1}$) of the plasmoids here are consistent with those in \citet{li16a}, but larger than those in \citet{2015ApJ...798L..11Y} and \citet{2020A&A...633A.121X}.
In NVST H$\alpha$ images, reconnection inflows toward the sheet-like feature are detected, with a mean speed of 6 km\,s$^{-1}$.
This inflowing speed is smaller than those in \citet{2015ApJ...798L..11Y} and \citet{2021ApJ...908..213L}.
The DEM-weighted temperature (1.7\,MK) of the current sheet is smaller than those in \citet{li16a, 2021ApJ...908..213L}.
The EM (1.1$\times$10$^{28}$ cm$^{-5}$) and electron number density (8.9$\times$10$^{9}$ cm$^{-3}$) are, however, larger than those in \citet{li16a, 2021ApJ...908..213L}. 
They are consistent with those during the reconnection process enhanced by the nearby filament eruption in \citet{2021ApJ...908..213L}.

\acknowledgments

The authors thank the referee for helpful comments that led to improvements in the manuscript. We are indebted to the SDO team for providing the data.
AIA images are the courtesy of NASA/SDO and the AIA, EVE, and HMI science teams.
This work is supported by the Strategic Priority Research Program of Chinese Academy of Sciences (CAS), grant No. XDB 41000000, the National Natural Science Foundations of China (12073042 and U2031109), the Key Research Program of Frontier Sciences (ZDBS-LY-SLH013) and the Key Programs (QYZDJ-SSW-SLH050) of CAS, and Yunnan Academician Workstation of Wang Jingxiu (No. 202005AF150025).
We acknowledge the usage of JHelioviewer software \citep{2017A&A...606A..10M}, and NASA's Astrophysics Data System.

\clearpage
\newpage

\appendix

\section{Formation and eruption of a flux rope}\label{sec:appa}

In the AR, besides the filament, we also identified a sigmoidal hot channel. Such a sigmoidal feature is generally considered to be a hot flux rope \citep{2012ApJ...761...62C, 2012NatCo...3..747Z, 2016ApJ...829L..33L}. We found that the feature forms above the PIL of the AR with the same magnetic connectivity as the filament, and it is clearly identified in AIA 94 and 131\,\AA~images; see Figure\,\ref{f:filament}(e). 

During the reconnection process between the filament and the emerging loops L2, it erupted outward from $\sim$04:08\,UT on April 22, with a mean projection speed of $\sim$80 km\,s$^{-1}$; see Figure\,\ref{f:filament}(h) and the online animated version of Figure\,\ref{f:filament}.
This eruption results in a C3.5 flare that peaks at $\sim$04:35 UT; see the flare ribbons and post-flare loops in Figure\,\ref{f:filament}(h) and the online animated version of Figure\,\ref{f:filament}, and a CME.
The filament remains stable under the post-flare loops during the flux rope eruption; see the online animated version of Figure\,\ref{f:filament}.
It is hence located underneath the hot flux rope, forming a double-decker configuration.
The reconnection between the filament and emerging loops L2 seems to be not affected by the flux rope eruption as well.
In NVST H$\alpha$ observations, the flare ribbons and post-flare loops of the C3.5 flare that are caused by the flux rope eruption are also detected; see Figure\,\ref{f:mr_nvst}(a).
Here, the post-flare loops show absorption features, representing the flare-driven coronal rain falling along the post-flare loops to the flare ribbons \citep{2021RAA....21..255L}.
Similar to the AIA EUV observations, the filament is located under the post-flare loops; see Figure\,\ref{f:mr_nvst}(a), supporting the double-decker configuration.

To investigate the formation of the flux rope, we check the evolution of EUV activities and magnetic fields along the flux rope before the eruption.
At the middle of the filament, and also the flux rope, repeated brightenings are observed; see Figures\,\ref{f:filament}(d), (f), and (g), and the online animated version of Figure\,\ref{f:filament}.
Plasmoids form in the brightening region, and propagate bidirectionally along the filament, with a mean speed of $\sim$150 km\,s$^{-1}$; see Figures\,\ref{f:filament}(d) and (g).
Underneath the brightening region, successive magnetic flux cancellation between the positive and negative magnetic fields P1a and N1a is identified; see the online animated version of Figure\,\ref{f:magnetic_fields}.
In the green rectangle in Figure\,\ref{f:magnetic_fields}(d), magnetic fluxes of the positive and negative magnetic fields, P1a and N1a, at the interface are calculated, and shown in Figure\,\ref{f:magnetic_fluxes}(a) as blue and green lines.
Both magnetic fluxes decrease significantly with a mean decreasing rate of $\sim$4$\times$10$^{18}$ Mx\,hr$^{-1}$, clearly indicating the magnetic flux cancellation; see Figure\,\ref{f:magnetic_fluxes}(a).
All the results point to magnetic reconnection at the middle of the filament (flux rope).
We suggest that reconnection between the loops, separately connecting the positive and negative magnetic fields P1 and N1a, and P1a and N1b, forms the longer higher-lying flux rope and the shorter lower-lying loops, connecting the positive and negative magnetic fields P1 and N1b, and P1a and N1a, respectively.
This is consistent with the formation mechanism of flux rope by tether-cutting reconnection \citep{2001ApJ...552..833M, 2014ApJ...797L..15C, 2016ApJ...823...41D, 2021ApJ...919L..21L}.
The submergence of the shorter lower-lying loops is followed by magnetic flux cancellation of the positive and negative magnetic fields P1a and N1a.

\section{Recurring EUV jets caused by emerging magnetic fields}\label{sec:appb}

Before the reconnection between the filament and its nearby emerging magnetic fields P3 and N3, positive and negative magnetic fields, P2 and N2, emerged in the context negative ones N1c from $\sim$13:52\,UT on April 21; see Figure\,\ref{f:magnetic_fields}(b).
After the emergence, they separated from each other; see Figure\,\ref{f:magnetic_fields}(c) and the online animated version of Figure\,\ref{f:magnetic_fields}.
Another negative magnetic field patch, N2a, also forms during this emerging process.
Different from the negative magnetic field N2, the negative ones N2a moved to the north after the emergence; see Figures\,\ref{f:magnetic_fields}(d)  and (e). 
New rising loops L1, connecting the emerging positive and negative magnetic fields P2 and N2a, are detected in GONG H$\alpha$ and AIA EUV images; see Figures\,\ref{f:filament}(a)-(d).
Similar to the loops L2, the loops L1 show dark structures, i.e., AFS, in H$\alpha$ images.
In AIA EUV observations, they are observed as dark structures in the early phase of their evolution, and then are seen as bright loops. 
EUV jet activity occurs repeatedly above the positive and negative magnetic fields P2, and N2 and N2a; see Figure\,\ref{f:filament}(b) and the online animated version of Figure\,\ref{f:filament}, with a mean projection speed of $\sim$130 km\,s$^{-1}$. 
At the endpoints of the jet, magnetic flux cancellation between the emerging positive magnetic field P2 and the preexisting negative ones N1c is identified, resulting in the disappearance of the former; see Figure\,\ref{f:magnetic_fields}(f).
The southern part of the emerging negative magnetic fields N2a cancelled with some small scale emerging magnetic fields, and the northern part merged with the context negative magnetic fields N1c; see the online animated version of Figure\,\ref{f:magnetic_fields}.
All the results suggest that reconnection occurs between the emerging magnetic fields, P2 and N2a (N2), and their overlying magnetic fields.
This reconnection results in the higher-lying recurring EUV jets and the lower-lying loops \citep[see a review in][and references therein]{2021RSPSA.47700217S}, which submerge and cause the magnetic flux cancellation between the positive and negative magnetic fields P2 and N1c.


\end{document}